\documentclass[twocolumn,prd,nofootinbib]{revtex4}
\usepackage{verbatim}
\usepackage{color}     
\usepackage{framed}
\usepackage{amsmath}
\usepackage{graphicx}  
\usepackage{tabularx}
\usepackage{wrapfig}   
\usepackage{hyperref}
\usepackage{booktabs}
\usepackage{subfigure}
\usepackage{xspace}

\newcommand\mchirp{$\mathcal{M}_c$}

\newcommand\ovrlp{\mathcal{O}_{ij}}

\begin{document}

\title{Supplementing rapid Bayesian parameter estimation schemes with adaptive grids}

\author{C. A. Rose}
\email{carose@uwm.edu}
\affiliation{Center for Gravitation, Cosmology, and Astrophysics, University of Wisconsin-Milwaukee, Milwaukee, WI 53201, USA}
\author{V. Valsan}
\email{vvalsan@uwm.edu}
\affiliation{Center for Gravitation, Cosmology, and Astrophysics, University of Wisconsin-Milwaukee, Milwaukee, WI 53201, USA}
\author{P. R. Brady}
\email{patrick@gravity.phys.uwm.edu}
\affiliation{Center for Gravitation, Cosmology, and Astrophysics, University of Wisconsin-Milwaukee, Milwaukee, WI 53201, USA}
\author{S. Walsh}
\email{wsinead@gmail.com}
\affiliation{Center for Gravitation, Cosmology, and Astrophysics, University of Wisconsin-Milwaukee, Milwaukee, WI 53201, USA}
\author{C. Pankow}
\affiliation{Center for Interdiscplinary Research and Exploration in Astrophysics (CIERA) and Department of Physics and Astronomy, Northwestern University, 2145 Sheridan Road, Evanston, IL 60208, USA}
\affiliation{Center for Gravitation, Cosmology, and Astrophysics, University of Wisconsin-Milwaukee, Milwaukee, WI 53201, USA}
\email{pankow@gravity.phys.uwm.edu}

\begin{abstract}
In the age of multi-messenger astrophysics, low-latency parameter estimation of gravitational-wave signals is essential for electromagnetic follow-up observations. In this paper, we present a new edition of the Bayesian parameter estimation scheme for compact binaries known as Rapid PE. Rapid PE parallelizes parameter estimation by fixing the intrinsic parameters (such as the masses and spins of the binary) to a grid around a search-identified intrinsic point and by integrating over the extrinsic parameters via Monte Carlo sampling. The original version of Rapid PE calculated an effective Fisher matrix to fix the intrinsic parameters to an ellipsoidal grid. Here we use rectilinear gridding in intrinsic space. The use of rectilinear intrinsic grids provides the option to more easily include spin and allows for adaptive grid refinements to mitigate search biases and map out the posterior more completely without sacrificing sampling efficiency. We demonstrate that this parameter estimation method produces reliable results for simulated gravitational-wave signals from binary neutron star mergers.
\end{abstract}
\maketitle

\section{Introduction}
\label{sec:intro}
The first discovery of a merging binary black hole via gravitational waves~\cite{PhysRevLett.116.061102} has heralded a new era of gravitational-wave astronomy. The most accurate measurements of these binaries come from Bayesian Markov Chain Monte Carlo (MCMC) and nested samplers~\cite{Veitch:2010,2011RvMP...83..943V,PhysRevD.91.042003,Ashton:2018jfp}, which map out the parameter space by comparing gravitational-wave emission models to data. MCMC and nested sampling parameter estimation methods have been very successful in a wide parameter space~\cite{LIGO-CBC-S6-PE,2011PhRvD..83h2002D,2011PhRvD..84f2003C,gr-extensions-tests-Europeans2011,Romero_Shaw_2020,gwastro-mergers-PE-Aylott-LIGOATest,2011ApJ...739...99N,gw-astro-PE-Raymond,PhysRevD.91.042003,TheLIGOScientific:2016wfe}, but can take a long time to converge. This situation is undesirable when prompt electromagnetic emission is expected to fade on the timescale of minutes~\cite{2012ApJ...746...48M,2014MNRAS.439..757G,2014MNRAS.437L...6K,2014MNRAS.437.1821M,2014ApJ...780...31T,2013ApJ...775...18B,first2years}.

Ref.\cite{PhysRevD.92.023002} introduced a rapid parameter estimation scheme, known as Rapid PE, which speeds up parameter estimation by fixing the intrinsic physical parameters of the binary to a grid and by marginalizing over other parameters with a non-Markovian Monte Carlo sampler. In this work we build upon this scheme, and therefore we will refer to ref.\cite{PhysRevD.92.023002} as the prequel. While both editions of Rapid PE center the intrinsic grid around a point identified by a low-latency gravitational-wave search, the main difference between the two pertains to the layout scheme of the intrinsic grid. In the prequel, the layout of the intrinsic grid was determined by an effective Fisher matrix, which was not flexible enough to overcome search biases~\cite{PhysRevD.77.042001} or handle a wide variety of potential events promptly enough. In this work, we continue to develop the layout scheme to make Rapid PE more flexible and extensible to a larger set of physical parameters, such as component spins. Replacing the radial layout from the effective Fisher matrix~\cite{gwastro-mergers-HeeSuk-FisherMatrixWithAmplitudeCorrections} with a rectilinear grid allows us to expand the intrinsic grid to higher-dimensional parameter spaces including spin, and provides a straightforward method to refine the grid without sacrificing sampling efficiency.

This work demonstrates that it is feasible to place a relatively small number of fixed grid points in intrinsic space to explore the region of posterior support. Continuing to incorporate pre-existing information from the search, we suggest a method to identify the preliminary region of the grid via precomputations of the inner-product space (overlap). At each point of this rectilinear initial grid, we can employ adaptive mesh refinement to place additional grid points in regions where the posterior has the most support, thus mapping out the posterior more efficiently. 

\subsection{Parameter Spaces}

As in the prequel, we subdivide the gravitational waveform parameters $\vec{\mu}=(\vec{\lambda},\vec{\theta})$ into two classes: the intrinsic parameters $\vec{\lambda}$ and the extrinsic parameters $\vec{\theta}$. Intrinsic parameters $\vec{\lambda}$ are intrinsic to the binary system itself, such as the masses and spins. In this work, we define various transformations of the binary component masses $m_1$ and $m_2$ (with $m_1>m_2$) as the coordinate space of the rectilinear intrinsic grid. One such mass combination is \mchirp ~and $\eta$, where \mchirp~$= (m_1 m_2)^{3/5}/(m_1 + m_2)^{1/5}$ is the chirp mass and $\eta = m_1 m_2/(m_1 + m_2)^{2}$ is the symmetric mass ratio. For comparison with the effective Fisher matrix grid of the prequel, we also place the grid in $(\tau_0,\tau_3)$ space with chirptime parameters $\tau_0 = 5(\pi M f_L)^{-5/3}/(256\pi f_L \eta)$ and $\tau_3 = (\pi M f_L)^{-2/3}/(8 f_L \eta)$, where $M = m_1 + m_2$ and $f_L$ is the lower cut-off frequency of the template~\cite{PhysRevD.76.102004}.

While the Fisher matrix scheme of the prequel was not extensible to spin, the rectilinear formulation of the grid in this work allows the option of gridding over the component spins of the binary $\vec{S}_1$ and $\vec{S}_2$ in addition to the mass parameters. Neglecting spin components can strongly influence the posterior position away from the true value~\cite{Farr:2015aa,Berry:2016oja}, especially when one or more components of the binary is a black hole~\cite{0004-637X-820-1-7}. 

When one or both of the components is a neutron star, finite size effects are most often encoded in the tidal deformability~\cite{2014arXiv1402.5156W}. We neglect tidal effects in this work as in the prequel, which for some equations of state, could impose futher biases~\cite{PhysRevD.91.043002}. We also assume that the binary has circularized before that portion of the inspiral is detectable by ground-based interferometers, and as a result, the eccentricity is negligible. 

At each intrinsic grid point, Rapid PE marginalizes over the extrinsic parameters $\vec{\theta}$, which primarily deal with the geometric orientation of the binary relative to the interferometric detector. The celestial position of the binary in the sky is given by the right ascension $\alpha$, the declination $\delta$, and the distance $D$. The orientation of the binary relative to the line of sight is encoded in the inclination angle $\iota$, the polarization angle $\psi$, and the phase $\phi$ at some reference frequency (usually taken to be the moment of coalescence). Finally, the geocentered time of coalescence $t_c$ establishes the epoch of the event. 

As demonstrated in the prequel, using pre-existing information from earlier searches provides a hierarchy in which each stage informs the next, increasing the efficiency of each subsequent step. Rapid PE reads the template producing the highest signal-to-noise ratio (SNR) from a low-latency gravitational wave search. The intrinsic point associated with that template determines the preliminary region of Rapid PE's initial grid based on the overlap. Rapid PE also reads in a posterior probability sky map from the low-latency sky localization routine BAYESTAR~\cite{gw-astro-Bayestar} to focus on the important regions of interest when sampling right ascension and declination. 

Gravitational-wave searches employ banks of templates~\cite{PhysRevD.76.102004}: a list of parameter sets used to generate template waveforms and match-filter the data. The product of the procedure is a list of candidate events corresponding to a set of parameters and a statistic reflecting the strength. The fineness of template banks is regulated with a minimal mismatch condition: within the bounds of the target space, any given template must have another template for which the mismatch does not exceed a certain fraction\footnote{This number is often $\sim 3$\%}. This implies that top level searches can quickly identify regions of interest for parameters included in the bank if the mismatch matrix derived from the template bank construction is available. Therefore, we can examine the inner-product manifold in the template coordinates to construct a neighborhood around a candidate by identifying all other templates within a desired overlap threshold. Another advantage of this concept is that any multi-modality in the posterior can be accounted for algorithmically.

\section{Motivation}
\label{sec:motivation}
In the prequel, the rapid parameter estimation scheme used information from a gravitational wave search to bound some parameters in its search. The scheme uses the epoch of the event, as well as any identified intrinsic parameters --- up to this point only a point estimate from the mass parameters. From this point estimate, the scheme then attempted to discern interesting bounds on the posterior integral. This region was constructed by using a local approximant of the ambiguity region. It inscribed an ellipse describing the approximate overlap at a given level around the search identified point in the \mchirp~and $\eta$ plane. It is not obvious how to efficiently cover this space, and rectilinear gridding is not straightforward. Moreover, multimodality in the space is not handled if only one event template point is reported. We turn to previously unused information from the gravitational wave searches to ameliorate these issues.

\subsubsection{Gravitational-Wave Search Template Banks}
\label{sec:template_bank_construction}

As of O3a, the first half of LIGO~\cite{LIGOScientific:2014pky} and Virgo's~\cite{VIRGO:2014yos} third observing run~\cite{Abbott_2021}, there are four low-latency gravitational-wave pipelines which use matched-filtering to identify gravitational-wave candidates based on banks of template waveforms~\cite{Cannon_2012,gstlal2017,Sachdev:2019vvd,Hanna:2019ezx,Allen:2005fk,Allen:2004gu,DalCanton:2014hxh,Usman:2015kfa,Nitz:2017svb,MBTA2016,MBTA2021,chu2021spiir}.  While Rapid PE can now draw information from any of these four modeled searches, for the purpose of this work we will focus on GstLAL~\cite{Cannon_2012,gstlal2017,Sachdev:2019vvd,Hanna:2019ezx}, the search pipeline used in the First Two Years (F2Y) injection campaign~\cite{first2years}. In the F2Y study, GstLAL imposed a single-detector signal-to-noise ratio (SNR) threshold of $\geq 4$ and a false alarm rate (FAR) of $\leq 10^{-2}$yr$^{-1}$ to recover low-spinning simulated binary neutron star (BNS) signals from a template bank of non-spinning TaylorF2 waveforms~\cite{Buonanno_2009}. Of the GstLAL-recovered candidates within these thresholds, only the event with the highest SNR is kept for each injected signal detected, similarly to how the ``preferred event"\footnote{\url{emfollow.docs.ligo.org/userguide/analysis/superevents.html\#preferred-event}} is chosen during an observing run~\cite{gwcelery}. Because this highest SNR template corresponds to the maximum log likelihood ratio ($\Lambda$)~\cite{Abbott_2020}, and thus the region of intrinsic space with the most posterior support, we use this search information to identify the region of Rapid PE's initial grid.

\begin{figure*}[htbp]
\includegraphics[width=0.99\textwidth]{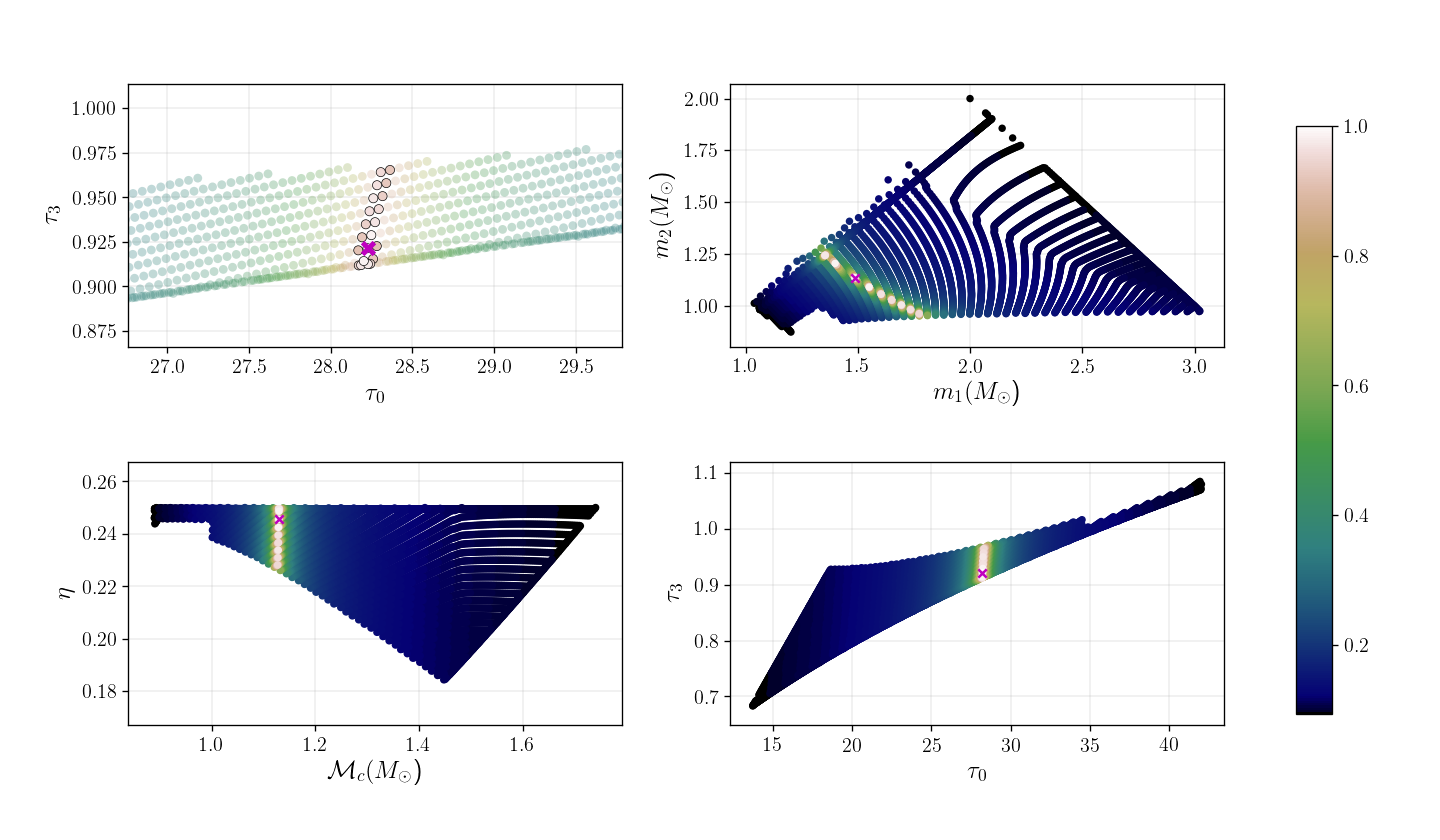}
\caption{\label{fig:bank_overlap} The template bank plotted with colorized overlap values for mass values corresponding to the event recovered from the GstLAL search with ID \#14631 from the prequel (1.49, 1.14 M${}_{\odot}$) --- here marked with a magenta cross. This bank was used in the F2Y paper to generate the initial candidate triggers which were then followed up by BAYESTAR and the Bayesian samplers. The top right panel is the overlap in component mass space. Points with overlap greater than 0.5 have been plotted more prominently to emphasize their position. The bottom left panel is in (\mchirp, $\eta$), and the bottom right is in ($\tau_0, \tau_3$) coordinates. With only mild ($|\vec{S}| < 0.005$) spin, the convex hull of overlap at a given threshold is particularly simple in $(\tau_0, \tau_3)$, suggesting its natural use as the coordinate basis. The top left panel shows a zoomed-in region with circumscribed points representing the templates with $>$90\% overlap.}
\end{figure*}

During a typical observing period, GstLAL's matched filtering process usually produces many more triggers which are either below the trigger production thresholds, rejected by signal consistency tests, or are clustered away~\cite{2012PhRvD..85l2006A,TheLIGOScientific:2016qqj}. This is because the template banks are produced at a fine resolution in a degenerate coordinate space. For example, a 3\% mismatch criteria means that an event with an SNR of 10 and a search trigger production threshold of 5.5 could produce a large number of triggers near where the match between templates is high. This reflects the degeneracy in the bank: the match of the templates which are ``near'' the maximal template translate to an SNR above threshold. For a visualization of this, see Figure \ref{fig:bank_overlap}. Each point in the template bank with a corresponding SNR measurement serves also as an approximation to the likelihood ratio at that point in parameter space.

Matched-filtering pipelines can therefore be considered as a gridded (though often much coarser and possibly biased) preliminary estimate of the posterior in the parameter space. In this work, we use the search template bank to identify the parameter space region of the signal. Typically the searches only report only the most significant template, although more detailed information can be made available. The identification of all templates which have a significant overlap with that template allows us to select the parameter space region with the highest likelihood. 

Once the region is identified, a grid is laid down on which to evaluate the reduced likelihood. In general, the template bank points are unsuitable since the fall off in the likelihood (and hence the posterior) is too steep --- it would be unlikely that more than a few points would have significant posterior support. At detectable SNRs, the likelihood function becomes very sharply peaked, easily overwhelming the prior and concentrating well within the spacing between parameter points in the bank. Instead, we use the template bank only for the initial identification, and choose a rectilinear grid in the parameter space for Rapid PE. The number of points in the initial region is chosen such that the space is adequately covered to find the maxima in the region. A more detailed set of measurements is then made by adaptive mesh refinement procedures. The mesh is refined only in areas where the posterior support is significant, thereby allowing us to explore the space in an efficient manner. The coupled approach of region identification and subsequent refinement allows for a controlled, efficient, and near-exhaustive search of the parameter space where there is pre-existing template coverage. 

\subsection{Preliminary Region Identification Method}

To enable efficient identification of the preliminary region, we precompute the overlap matrix. For each template in the bank, we calculate the overlap inner-product for each other template in the bank, producing a symmetric matrix of values between 0 and 1. As in ref.\cite{PhysRevD.92.023002}, the overlap is given by 

\begin{equation}
\label{eqn:ovrlp_heur}
\ovrlp = \frac{|\langle h_i|h_j \rangle|}{\sqrt{\langle h_i|h_i \rangle \langle h_j|h_j \rangle}}
\end{equation}
where 
\begin{equation}
\label{eqn:inner_prod_def}
\langle a | b \rangle = 2 \int_{-\infty}^{\infty} \frac{\tilde{a}(f) \tilde{b}(f)^{\ast}}{S(f)} df\qquad.
\end{equation}
is the noise-weighted inner product of two time-series $a(t)$ and $b(t)$, with weighting function $S(f)$ the noise power spectral density. Neighborhoods are constructed by setting the gravitational-wave signal $h_i$ to the template indicated by the search, and selecting the set of $h_j$ such that $\mathcal{O}_{ij}$ is greater than a given threshold value. 

The overlap represents the similarity of two waveforms. It is a characterization of closeness, and is related to the expected SNR $\bar{\rho}$ of an event, given by $\bar{\rho}^2=\langle h | h \rangle$, where $h$ is the signal strain.

The region formed from the effective Fisher matrix in the prequel~\cite{PhysRevD.92.023002} is replaced with a region which circumscribes the neighborhood of points around a given template identification. These regions more accurately identify the ambiguity with the input template up to the granularity of the template bank itself. Neighborhoods are not required to be concentrated near the input point in the coordinate distance sense. Compact neighborhoods will lend themselves well to refinement procedures (outlined in section \ref{sec:refinement}), requiring fewer refinement levels to accurately capture the posterior mass. 

The overlap threshold sets the minimum overlap between the search-reported point and the set of points in the precomputed overlap matrix used to determine the initial region of the intrinsic grid. For our examples in section \ref{sec:intr_evidence}, we choose an overlap threshold of 0.97 based on the fineness of the template bank used by GstLAL. We demonstrate that determining the overlap threshold based on the template bank fineness alone does not always cover the true parameters of the signal due to search biases such as neglecting spin. Therefore in the case of significant search biases, it is beneficial to choose a lower overlap threshold than the match of the templates in the bank, as demonstrated in section \ref{sec:validation}. 

\subsubsection{Grid Setup and Refinement}

Once the region is identified, a gridding scheme can be applied. While the grid could be constructed in any applicable coordinate space, given our choice of a regular grid, it is most efficient to do so in a space where the shape of the region selected is closest to rectilinear. From template bank studies~\cite{Babak_2013,PhysRevD.92.023002}, it is known that the region created in (\mchirp, $\eta$) is elliptical, with the major axes rotated relative to the coordinate axes. Since the template bank used in this work is constructed in the $(\tau_0, \tau_3)$ space~\cite{first2years,Babak_2013}, and the templates themselves will have a fixed spacing here, we choose these parameters to construct our grid. 

We place a grid of $N_p\times N_p$ points along the coordinate axes. As shown in Fig. \ref{fig:bank_overlap}, even in the $(\tau_0, \tau_3)$ space this grid has a non-negligible number of points outside the preliminary region. To better capture the shape of the posterior support with efficient computations, we recalculate the overlap over the initial grid and deactivate all grid points which do not meet the overlap threshold criteria used to create the preliminary region. The remaining grid represents the fixed intrinsic parameters passed to the extrinsic integrator presented in the prequel.

\section{Mesh Refinement Strategies}
\label{sec:refinement}
A fixed grid will only capture the peak of the distribution to a resolution of the grid. The na\"ive solution to this problem is to increase $N_p$, the initial number of grid points on a side. The deactivation strategy above can ameliorate some wasted computations. If instead, we employ adaptive mesh refinement (AMR)~\cite{1989JCoPh..82...64B}, we can concentrate points where the posterior has the most support and increase the number of effective samples collected in the integration process. AMR is used heavily in numerical relativity and computational astrophysics~\cite{Dubey20143217}, where precise control over resolution is required to accurately represent the spacetime around singularities or efficiently represent quantities with complex dynamics and dynamic ranges. We face a similar situation where the peak of the posterior scales as $\exp(+\bar{\rho}^2)$ where $\bar{\rho}$ is the expected signal to noise ratio. Even small mismatches between templates on the grid and a template at the true signal parameters can cause a significant underestimate of the peak for typical SNR events. 

\begin{figure}[htbp]
\includegraphics[width=0.99\columnwidth]{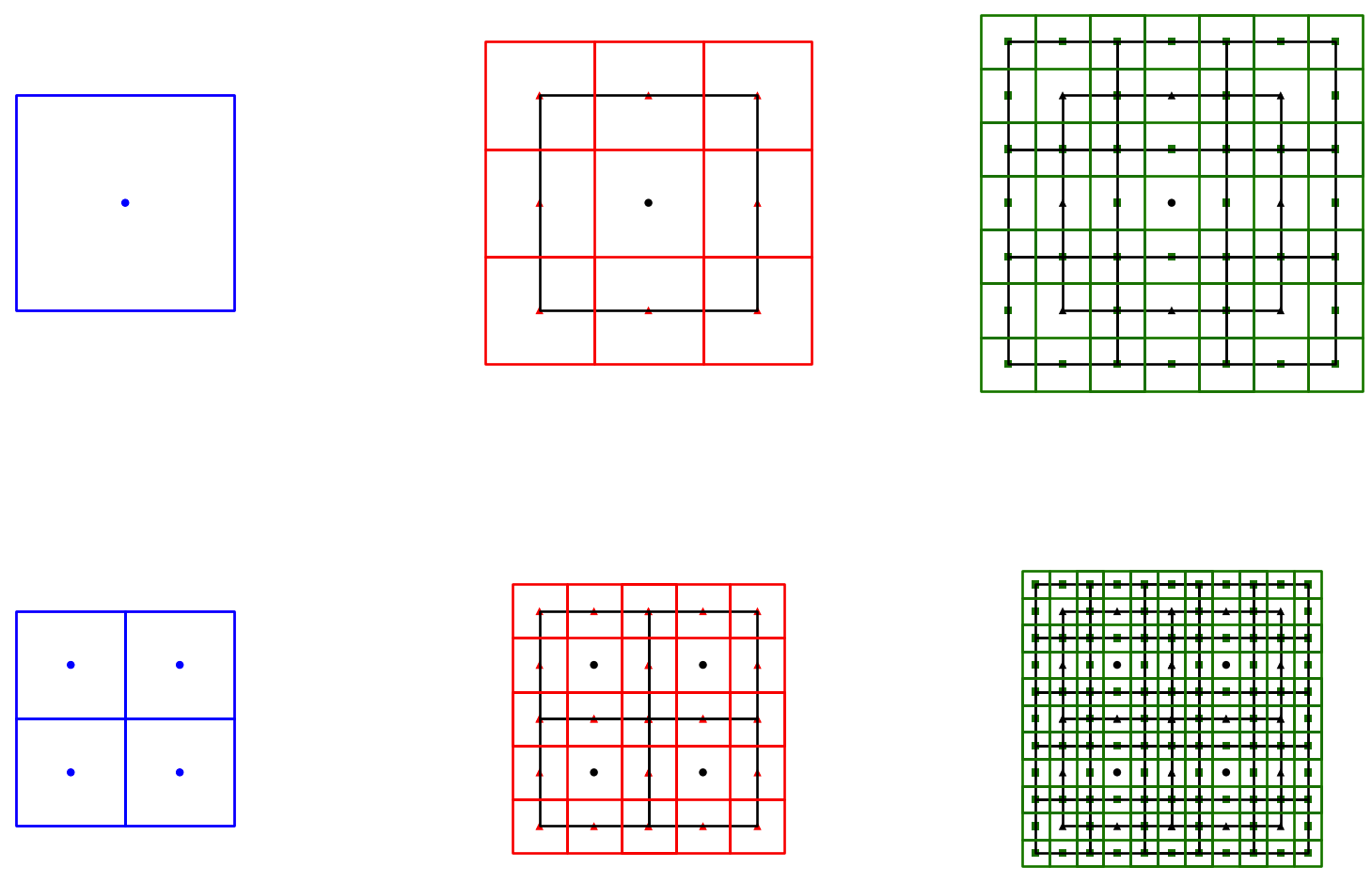}
\caption{\label{fig:grid_refine} The top three panels represent a fiducial cell and inset grid point (blue), a single level of refinement (red), and a second level of refinement (green). The bottom three panels display the same procuedre, but with a split initial region to highlight the interlocked but overlapping refined cells. In all cases, the black cells are the previous level underlayed for reference. Grid points which have already been evaluated (the center point, zero offset refined cell) are not included in the next refinement level.}
\end{figure}

Ideally, the refined subgrid resolution should scale with the value of the posterior itself. Moreover, a careful examination of the correlations on a point by point basis (e.g. a princical component analysis transformation) might lead to a more ideal refinement scheme. However, we opted instead for a simpler refinement method which produces points at the midpoints between grid points and their diagonals --- see Figure \ref{fig:grid_refine}. First, the points to be refined are selected by the criteria outlined later in the section. Then, each grid point to be refined is considered to be at the center of a cell extending to the midpoint between the grid point neighbors, so the new subgrid points will lie along the edges and vertices of the cell. Explicitly, the subgrid is constructed for each point in the parent grid by iterating over all possible permutations of cell displacement and 0 vectors in the $d-$dimensional space with the displacement vectors pointing along the coordinate axes, here indexed as $a_{0,1,\cdots d}$. The cell displacement vectors have a magnitude of half the parent cell side length. In two dimensions this corresponds to nine non-degenerate offsets: $\{0, 0\}, \{0,\pm a_2\}, \{\pm a_1,0\}, \{\pm a_1,\pm a_2\}$. The zero displacement vector is omitted since it corresponds to a grid point which has already been computed. Additionally, adjacent cells on the level to be refined will produce offset vectors which point to the same grid point (e.g. the bisector between two grid points). Once the refinement procedure is complete, duplicate cells are removed.

The initial choice of $N_p=5$ is typically coarse enough to eliminate large regions where the overlap is below threshold. Five points per dimension is too few to effectively locate the peak and explore the support of the posterior, even in two dimensions. So, before likelihood evaluation on the initial grid, we apply one level of refinement to the grid. This procedure can be repeated as many times as is prudent or computationally feasible, potentially producing several subgrid evaluation levels.

We now elaborate on two possible refinement protocols:

\begin{itemize}
\item \emph{serial refinement}: Where the grid refinement is based on the calculated value of the reduced likelihood at the grid point
\item \emph{prerefinement}: Where the grid refinement is based on the overlaps calculated on the grid points before any reduced likelihood computations are done
\end{itemize}

The latter allows one to place an arbitrarily refined grid in advance, allowing all reduced likelihood calculations to proceed in parallel. The former is still parallel, but each refinement step must wait for the previous set of reduced likelihoods to be calculated first.

\subsubsection{Serial Refinement}

Without modification, this procedure will produce a new grid with (at most) $3^d-1$ points around each grid point, where $d$ is the number of dimensions in the intrinsic space. However, the evaluation of the initial grid has provided additional information about the shape of the posterior. We can select the grid regions which need refining by confining our attention to the convex hull which contains 90\% (for example) of the probability mass as measured from the posterior evaluations already completed. In the examples shown, for binaries without strong spins, this is usually only a small fraction of the original set. 

In the case of serial refinement, a ``pruning'' step is applied after the gridding. This pruning calculates the overlap of each point in the grid with the input search point, and discards points with values less than a loose ad-hoc threshold. The examples presented in section \ref{sec:intr_evidence} use a loose threshold of 0.9.

\subsubsection{Prerefinement}\label{ssec:prerefine} 

In this case, a set of overlap thresholds are fixed in advance, correponding to various refinement levels. Subsequent levels are identified by recalculating the overlap with the initial point for each grid point, and those exceeding the predetermined overlap threshold are kept and used for the next level. This procedure can be repeated as many times as necessary.

\section{Intrinsic Evidence}
\label{sec:intr_evidence}
The following section illustrates Rapid PE results for two injections from the F2Y mock data challenge~\cite{first2years,first2yearsweb}. This dataset simulated the performance of the LIGO-Virgo interferometer network in a realistic configuration during 2015 and 2016 on injected BNS signals with mild, isotropic spins. In the F2Y study, the search pipeline used a non-spinning template bank and only provided estimates for the masses of the injections, so we will neglect spin as well for these examples. This induces biases~\cite{Farr:2015aa,Berry:2016oja} but does not hamper posterior recovery severely for the low-spin BNS injections discussed here. 

For comparison with the Fisher matrix grid of the prequel, we construct the rectilinear grid in $(\tau_0,\tau_3)$ coordinate space and plot it in (\mchirp,~$\eta$) space, producing a grid which is closely aligned with the primary axes of the ambiguity ellipse. The overlaps for the points within the ellipse produce values consistent with the 97\% contour measured by the effective Fisher matrix. 

We compare the three intrinsic grid methods described above, here denoted as ``Fisher" for the Fisher matrix approximation scheme of the prequel, ``grid refine" for the iterative refinement procedure using reduced likelihood calculations, and ``prerefine" for the single step refinement procedure using overlaps. 

\subsection{Event 10184}
\begin{figure*}[htbp]
\centering
\includegraphics[width=0.95\textwidth]{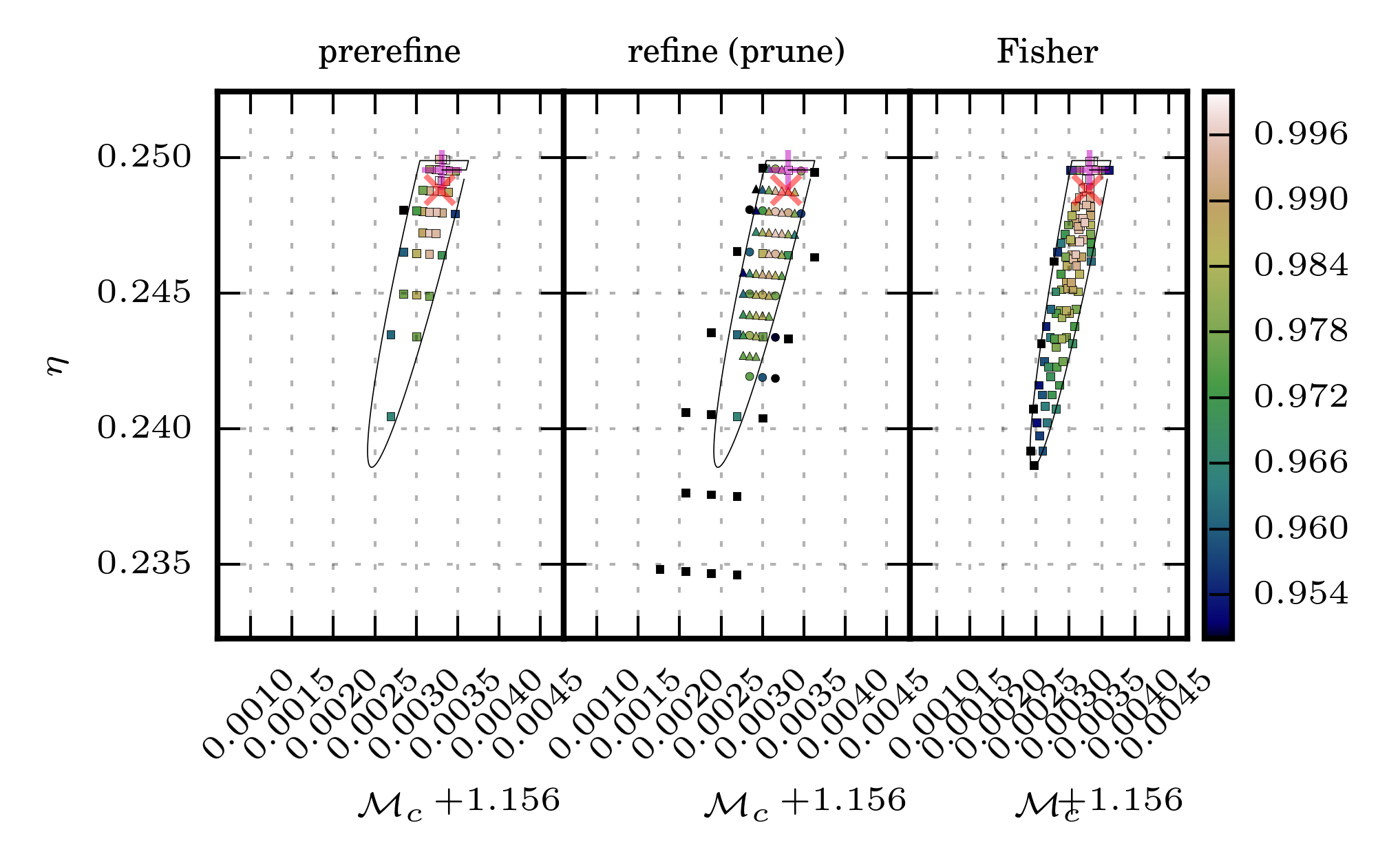}
\caption{\label{fig:grid_compare_intr_olap_10184} This figure shows the value of the overlap calculated for each sample intrinsic point scattered across the \mchirp~and $\eta$ plane for event \#10184. The left panel is the result for the prerefinement method, the middle is the (pruned) serial grid refinement, and the right the is Fisher matrix based method. The magenta cross marks the search-identified point from which the intrinsic grid was computed, while the red x marks the injected parameters. In all panels, the 97\% overlap contour obtained from the effective Fisher matrix is marked by a black line for comparison.}

\includegraphics[width=0.95\textwidth]{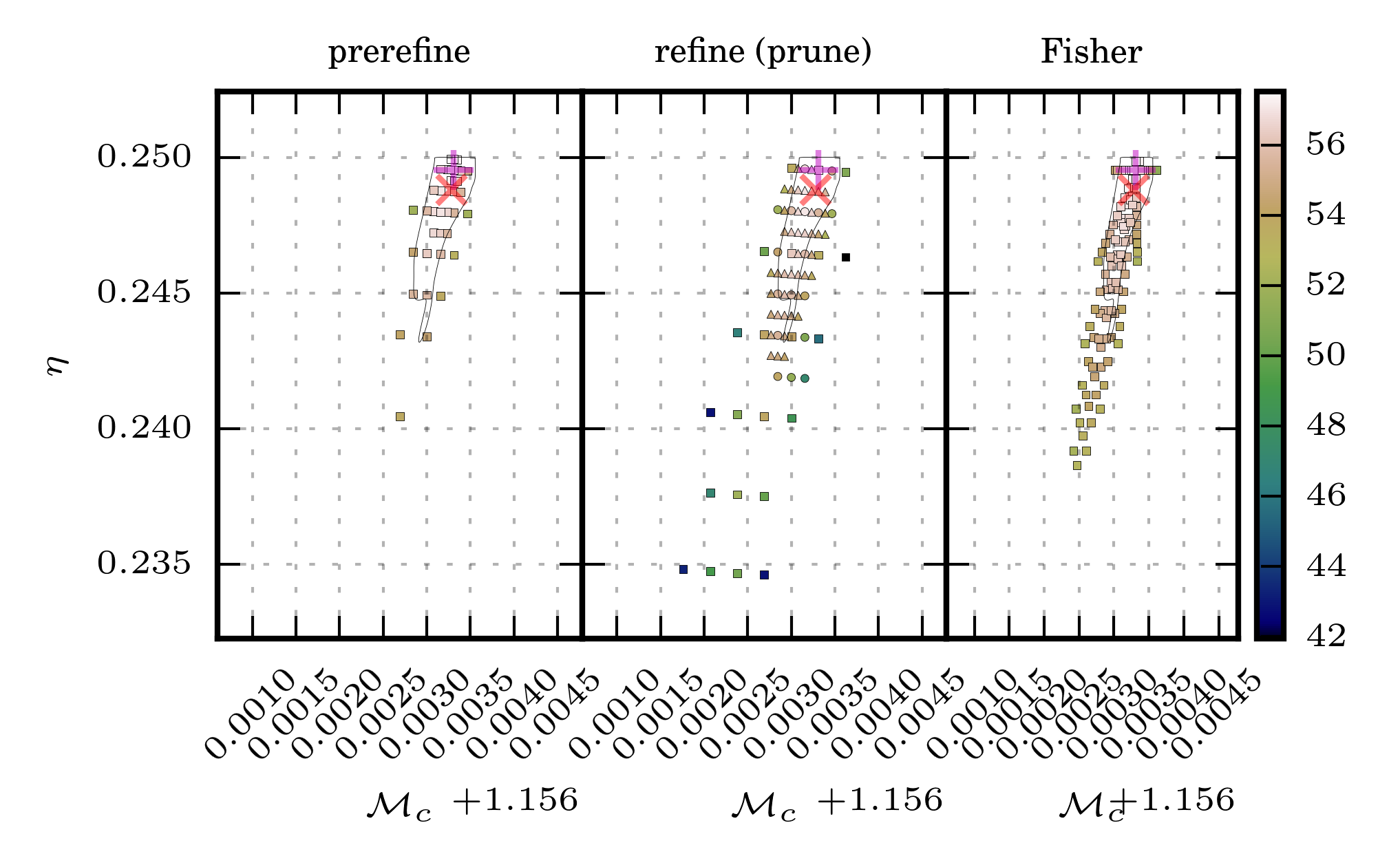}
\caption{\label{fig:grid_compare_intr_evid_10184} Here the logarithm of the reduced likelihood obtained through marginalization is scatter plotted against \mchirp~and $\eta$ for each method (prerefinement, pruned refinement, and Fisher based) for event \#10184. The magenta cross marks the search-identified point from which the intrinsic grid was computed, while the red x marks the injected parameters. In this figure, the contour represents the 90\% credible interval obtained from \texttt{lalinference\_mcmc}.}
\end{figure*}

\begin{figure*}[htbp]
\includegraphics[width=0.95\textwidth]{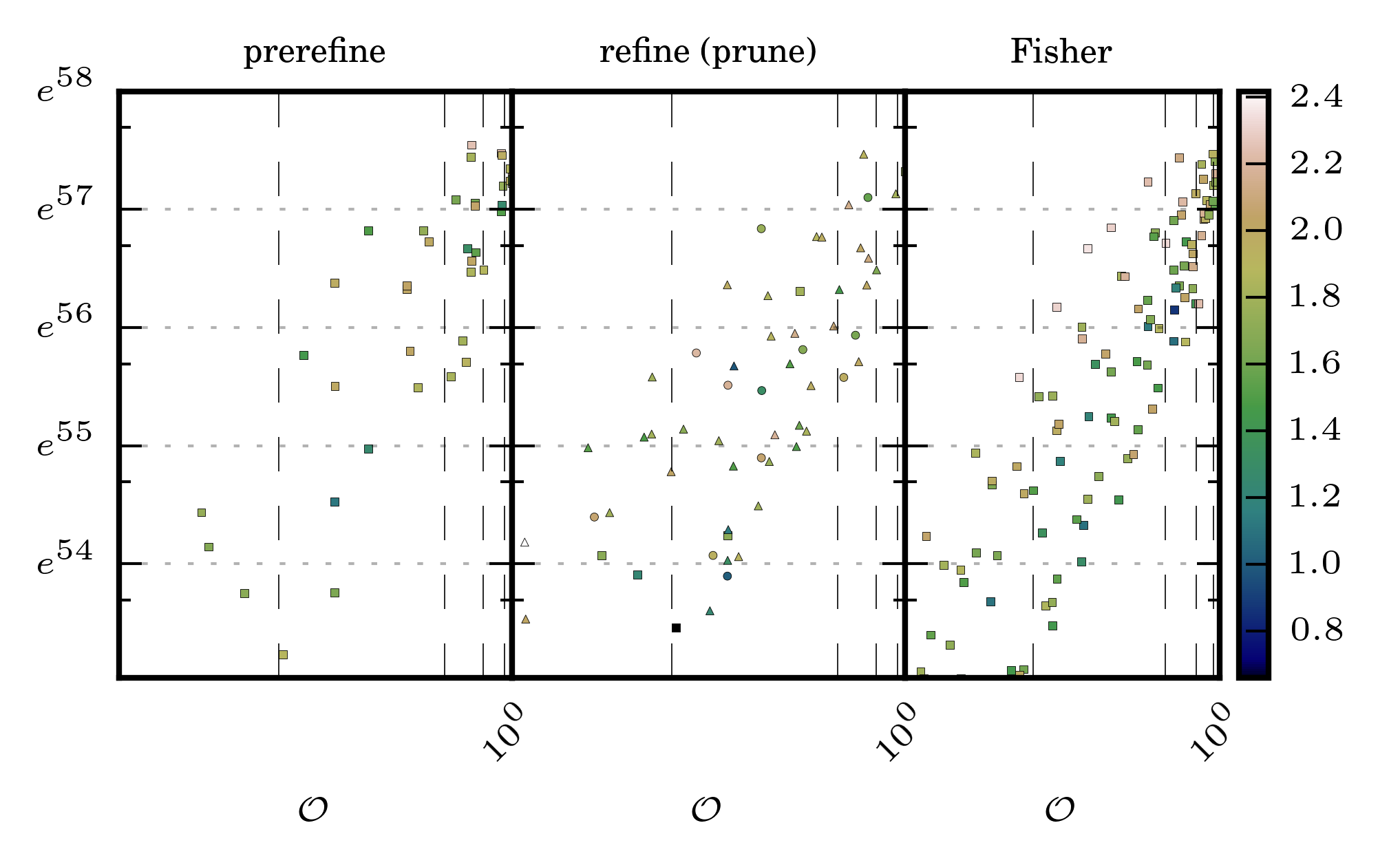}
\caption{\label{fig:grid_compare_evid_10184} For the three methods presented, the reduced likelihood is scatter plotted against the value of the overlap for that intrinsic point against the search value for event \#10184. The color scale indicates the base 10 logarithm of the number of effective samples collected for the extrinsic integration at that point.}
\end{figure*}

\begin{figure*}[htbp]
\includegraphics[width=0.95\textwidth]{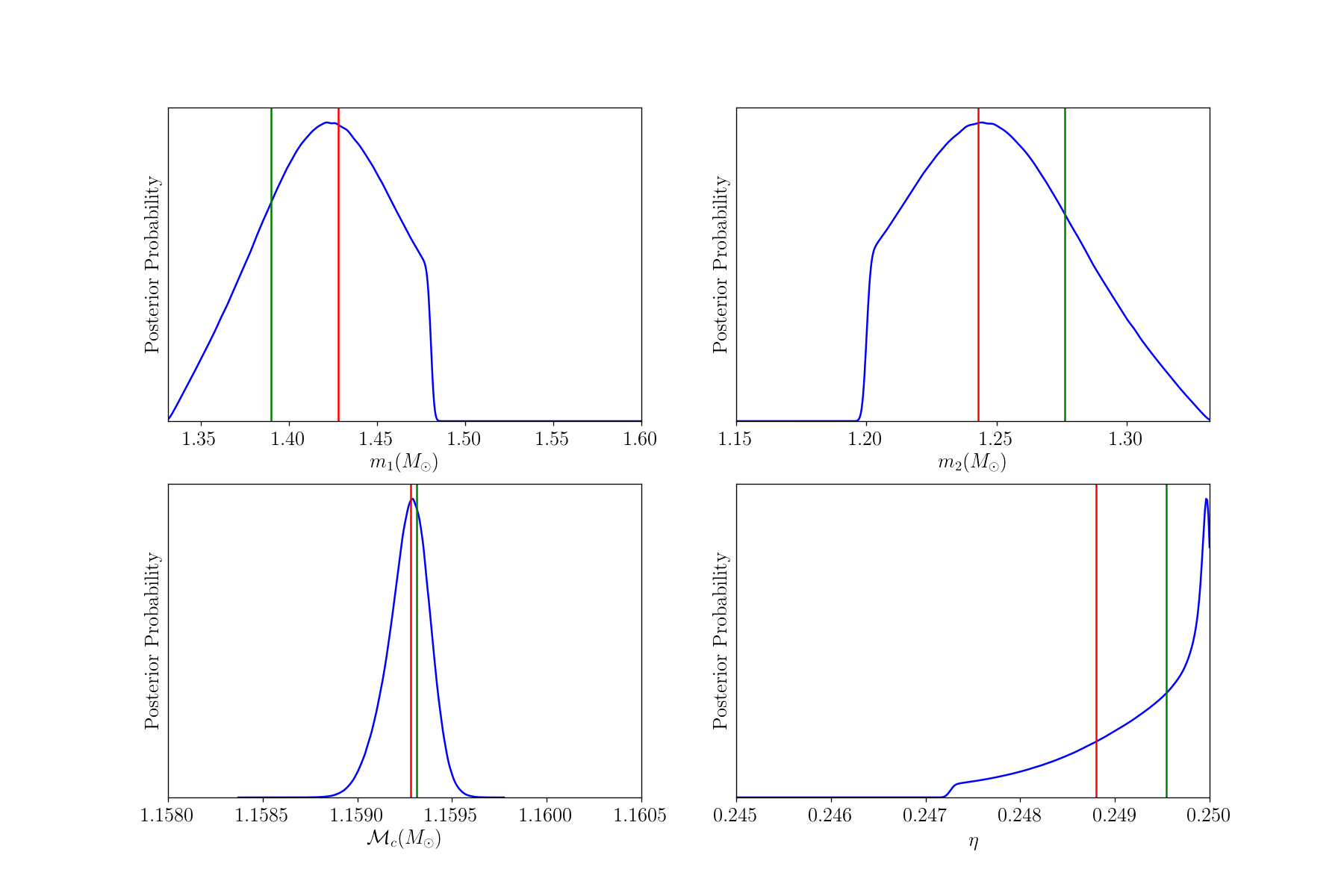}
\caption{\label{fig:posterior_10184} Intrinsic posterior plots of event \#10184 for the intrinsic grid based on overlaps between the search-recovered mass parameters. The red lines indicate the injected values, while the green lines indicate the search-recovered values.}
\end{figure*}

For the first example, we select an injection from the F2Y dataset with the search ID \#10184, network SNR of 13.6, search-recovered masses of $m_1=1.39\textup{M}_\odot$ and $m_2=1.28\textup{M}_\odot$, injected masses of $m_1=1.43\textup{M}_\odot$ and $m_2=1.24\textup{M}_\odot$, and dimensionless angular momentum aligned spin $\chi_a = -0.009$ and in-plane spin $\chi_p = 0.030$. Since the searches in the F2Y study do not consider spin, only the injected spins are listed. 

Figure \ref{fig:grid_compare_intr_olap_10184} shows the overlap computed against the search identified intrinsic point over all selected intrinsic points for each of the three methods. In all cases, the points which would have unphysical $\eta$ ($>0.25$) have been omitted. The right panel of Figure \ref{fig:grid_compare_intr_olap_10184} shows the effective Fisher matrix placement scheme, with points placed along radiating lines from the search point. In the case of event \#10184, the search identified point (magenta cross) is not greatly biased away from the injected location. The values of the overlap with the search identified point track correctly with the expected fall-off of the overlap.

The grid-refine method with pruning (middle column) shows an initial grid region corresponding roughly to a convex hull containing points with overlap $>0.9$. The grid points chosen for refinement $R^{(0)}\rightarrow R^{(1)}$ correspond to the top 90\% of the total marginalized likelihood for each level. 

The prerefinement method is shown in the left panel of Figure \ref{fig:grid_compare_intr_olap_10184}. Here the density of points is much higher than in the iterative grid refinement procedure because the prerefinement has an additional refinement level based on overlap recalculations as described in Section \ref{ssec:prerefine}.

Figure \ref{fig:grid_compare_intr_evid_10184} shows the value of the logarithm of the reduced likelihood after the Monte-Carlo integration step has completed for all three cases. The contour in this figure represents the 90\% credible region obtained from \texttt{lalinference\_mcmc}~\cite{lalsuite}. The high likelihood support in this region shows that all three methods perform reasonably well when search biases are negligible. 

We obtain intrinsic posterior samples from the Rapid PE grids by approximating the likelihood function as a sum of gaussians centered at each grid point with standard deviations of half the spacing between adjacent grid points, and weighted by the value of the extrinsic-marginalized likelihood calculated at each grid point. Figure \ref{fig:posterior_10184} shows the posterior plots for event \#10184, which are the histograms of the likelihood samples weighted by the mass priors. 

Finally, Figure \ref{fig:grid_compare_evid_10184} shows the marginalized likelihood as a function of the computed overlap. The trend of the increasing log likelihood versus overlap is clear, again reinforcing that using the overlap as a proxy for the likelihood is justified when biases are expected to be small.

\subsection{Event 14631}
\begin{figure*}[htbp]
\includegraphics[width=0.99\textwidth]{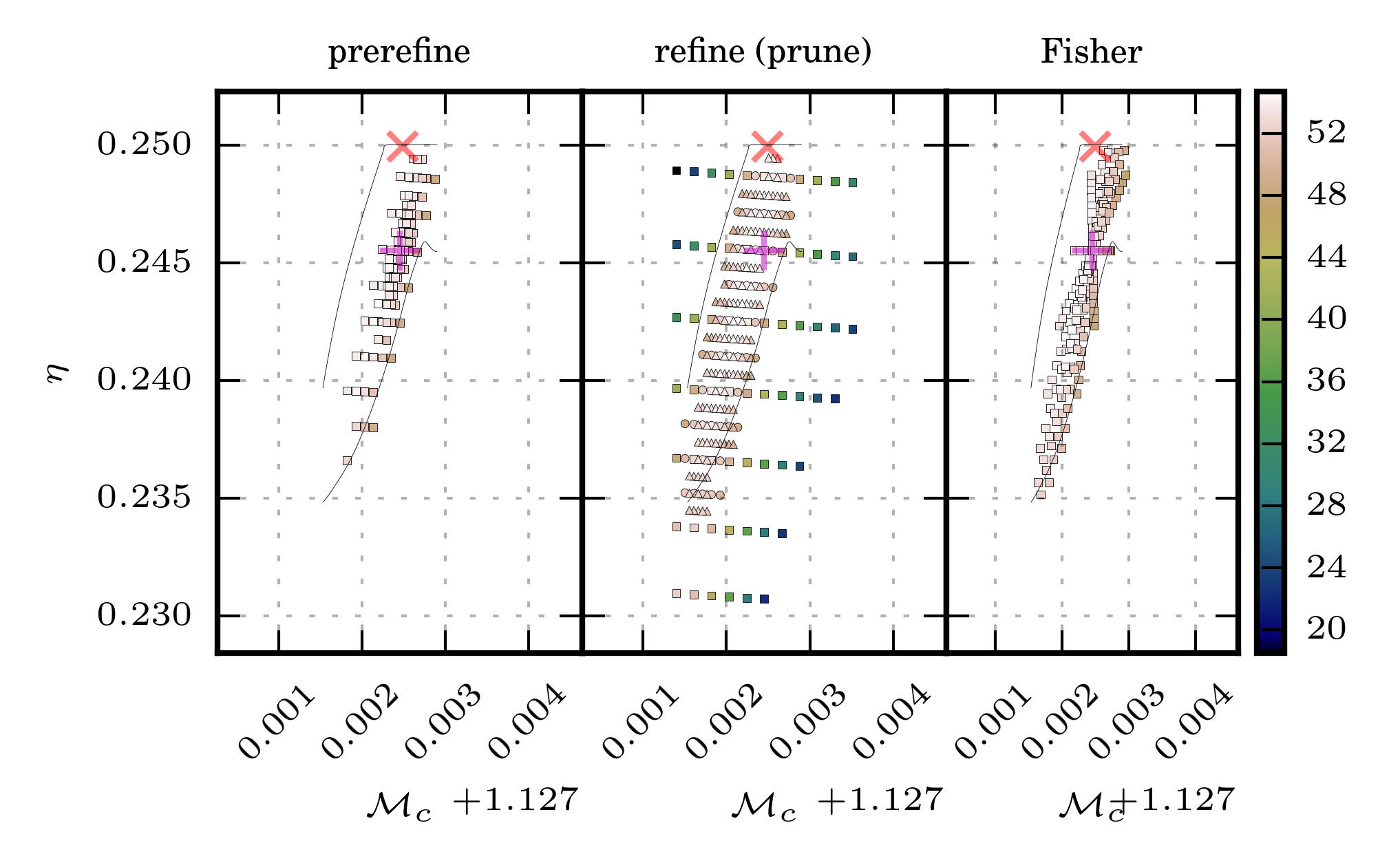}
\caption{\label{fig:grid_compare_intr_olap_14631} This figure shows the logarithm of the reduced likelihood obtained through marginalization scatter plotted against \mchirp~and $\eta$ for each method. The left panel is the result for the prerefinement method, the middle is the (pruned) serial grid refinement method, and the right the is Fisher matrix method. The magenta cross marks the search-identified point from which the intrinsic grid was computed, while the red x marks the injected parameters. In this figure, the contour represents the posterior obtained from \texttt{lalinference\_mcmc}.}
\end{figure*}

\begin{figure*}[htbp]
\includegraphics[width=0.95\textwidth]{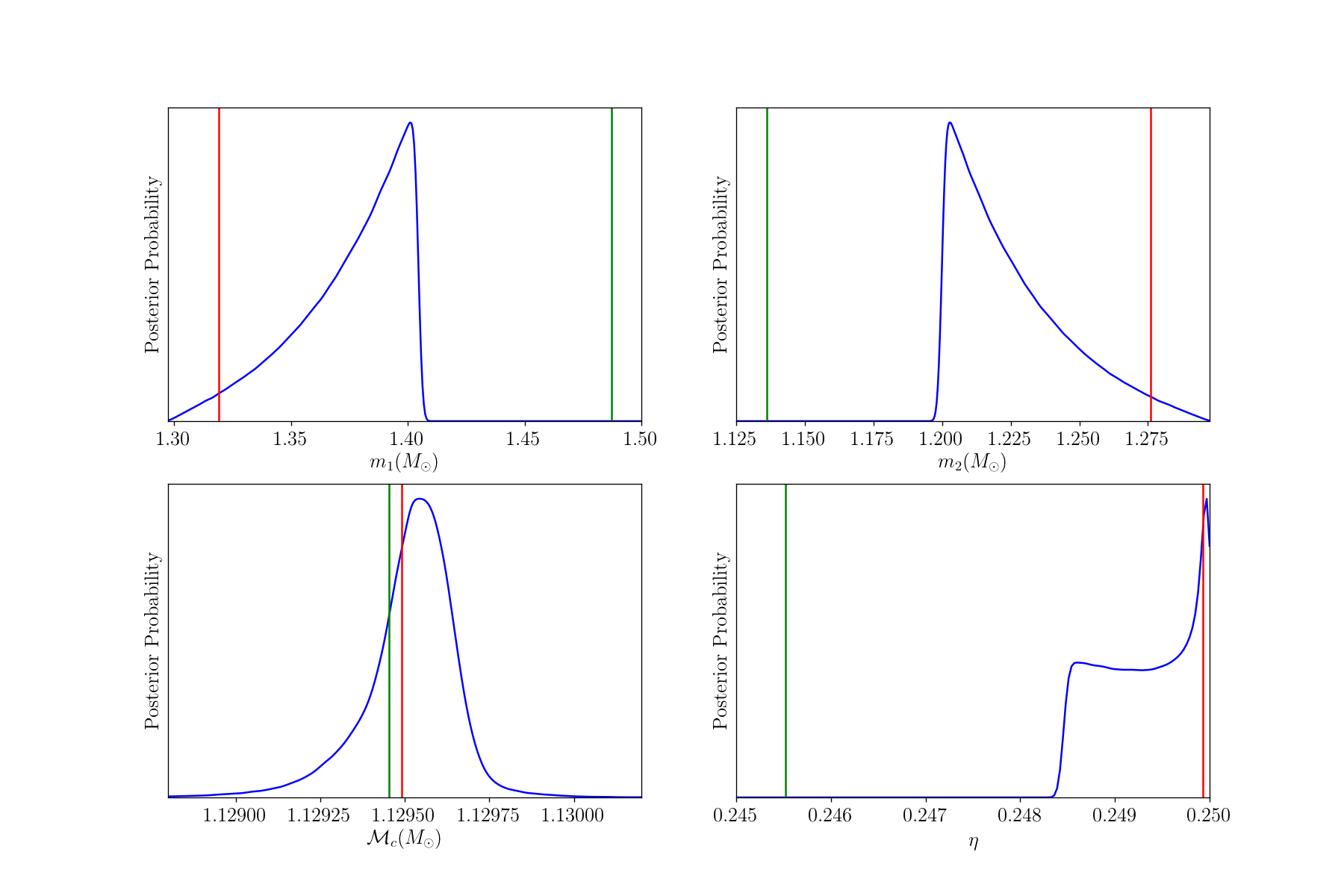}
\caption{\label{fig:posterior_14631} Intrinsic posterior plots of event \#14631 for three grid levels of the refine (prune) scheme. The red lines indicate the injected values, while the green lines indicate the search-recovered values.}
\end{figure*}

\begin{table}[htbp]
\begin{tabular}{cccc}
\hline
method & $N_{\text{eff}}$ & $N$ & ratio \\
\hline
\hline
refine & 832 & $2.5\times 10^{8}$ & $3.4\times 10^{-6}$ \\
prerefine & 468 & $1.3\times 10^{8}$ & $3.5\times 10^{-6}$ \\
Fisher & 424 & $1.7\times 10^{8}$ & $2.5\times 10^{-6}$ \\
\hline
\end{tabular}
\caption{\label{tbl:neff_n} This table displays the total number of effective samples $N_{\text{eff}}$, the total number of samples taken $N$, and the ratio for each of the three grid methods executed on event \#14631. The ratio serves as a measurement of the sampling efficiency of the method.}
\end{table}

 This example illustrates how the three grid methods perform for an event with more pronounced search biases. This injection from the F2Y dataset has a network SNR of 12.0, search-recovered masses of $m_1=1.49\textup{M}_\odot$ and $m_2=1.14\textup{M}_\odot$, injected masses of $m_1=1.32\textup{M}_\odot$ and $m_2=1.28\textup{M}_\odot$, and injected $\chi_a = 0.006$ and $\chi_p = 0.032$. Due to the search bias present in the detection of this event, the true parameters of the injection lie just on the edge of the 97\% contours as shown in Figure \ref{fig:grid_compare_intr_olap_14631}. The grid refinement scheme contains it, but it is just outside of the prerefine and Fisher schemes. 

In the Fisher and prerefine cases, where serial refinement is not available to compensate, the coverage of the total posterior is incomplete, as shown by the increasing reduced likelihood gradient towards decreasing \mchirp (see Figure \ref{fig:grid_compare_intr_olap_14631}, right and left panels). The grid refine method covers the full width of the posterior in \mchirp, showing that incompleteness is present but likely would not drastically affect an interpolated result. The serial refinement of this method sets up a wide enough grid that the \texttt{lalinference\_mcmc} contour is completely enclosed initially and subsequent refinements are well contained by the contour. Qualitative comparisons of the posteriors obtained by \texttt{lalinference\_mcmc} and the grid refinement method are favorable. 

Figure \ref{fig:posterior_14631} illustrates how even when the initial Rapid PE grid based on the search estimate completely misses the true injected intrinsic point, the adaptive grid refinement scheme of Rapid PE compensates for search biases well enough to resolve the peak of the posterior, which is much closer to the injected parameters than the search-estimated values. 

Numerically, the number of effective samples $N_{eff}$ provides a practical measure of how reliable we expect our posterior to be~\cite{PhysRevD.92.023002}. The efficacy of each method to collect effective samples $N_{eff}$ for this event is quantified in Table \ref{tbl:neff_n}. The ratio of the number of effective samples computed overall to the number of samples computed total gives an indication of how many samples need to be computed to accurately measure the shape of the posterior. Therefore $N_{eff}/N$ serves as a measurement of the sampling efficiency of each method. For event \#14631, this ratio is about the same for all three grid schemes, while the refine method has nearly double the $N_{eff}$ compared to the prerefine and Fisher methods. 

This event demonstrates how the adaptability of the serial grid refinement method overcomes search biases better than the Fisher or prerefine methods without sacrificing sampling efficiency. It also highlights the dangers involved in relying on a biased search result, and indicates a need to account for search biases with a wider grid. We take these lessons and apply them to our large scale test of Rapid PE in section V.

\section{Validation of Rapid PE}
\label{sec:validation}
To test the validity of Rapid PE, we ran it on 100 of the three-detector simulated signals from the F2Y mock data challenge injected into gaussian noise (via \texttt{gstlal\_fake\_frames}\footnote{\url{lscsoft.docs.ligo.org/gstlal/gstlal/bin/gstlal_fake_frames.html\#gstlal-fake-frames}}) following the mid PSD curve as in the 2016 study~\cite{first2years}. The initial grids were generated rectilinearly in $\mathcal{M}_c,\eta$ with 5 points per side and an overlap of 0.3 with the injected masses. We used the serial refinement method with a total of three grid levels, including the initial grid. The likelihoods were evaluated using the TaylorF2 waveform~\cite{Buonanno_2009} as in Section IV. We used the results of these runs to calculate the CDF of the posterior up to the injected parameter and check that the correct fraction of events are found within the given probability interval, or confidence interval (C.I.). For gravitational-wave parameter estimation codes, this is traditionally presented as a P-P plot, where the P's could stand for either probability or percent~\cite{Romero_Shaw_2020,PhysRevD.91.042003,Biwer_2019}. 

The First Two Years injection set includes component masses uniformly distributed from $m_L=1.2 \textup{M}_\odot$ to $m_H=1.6 \textup{M}_\odot$. Therefore the CDF of the posterior over $m_1$ for each injection is
\begin{equation}
        \mathrm{CDF}[\hat{m}_1]=\int^{\hat{m}_1}_{m_L}{dm_1} \int^{m_H}_{m_L}{dm_2 \: p(m_1,m_2 | d)}
\end{equation}
where $\hat{m}_1$ is the injected value of $m_1$ and $p(m_1,m_2)$ is the posterior distribution. 

By Bayes theorem, the posterior distribution is
\begin{equation}
        p(m_1,m_2 | d) = \mathcal{L}(m_1,m_2 | d)\:p(m_1,m_2)/\mathcal{Z}
\end{equation}
where $ \mathcal{L}(m_1,m_2 | d)$ is the likelihood function of the data, $p(m_1,m_2)$ is the prior distribution, and $\mathcal{Z}$ is the evidence
\begin{equation}
        \mathcal{Z}=\int^{m_H}_{m_L}{dm_1} \int^{m_H}_{m_L}{dm_2 \: \mathcal{L}(m_1,m_2 | d)\:p(m_1,m_2)}
\end{equation}

Rapid PE implicitly requires $m_1 > m_2$, so our mass prior distribution is
\begin{equation}
        p(m_1,m_2) = \frac{\Theta(m_1 - m_2)}{(m_H - m_L)^2} 
\end{equation}
where $\Theta(m_1 - m_2)$ is the heaviside step function.

At each intrinsic grid point, Rapid PE calculates the likelihood marginalized over the extrinsic parameters $\mathcal{L}_i$. To interpolate the marginalized likelihood in intrinsic space, we assume that the marginalized likelihood function is a sum of two-dimensional Gaussians in $\mathcal{M}_c-\eta$ space centered at each grid point with an amplitude of $\mathcal{L}_i$ and a standard deviation of half the grid spacing for each intrinsic parameter. We can approximate the likelihood function as,
\begin{multline}
        \mathcal{L}(m_1,m_2 | d)\approx \sum_i \mathcal{L}_i \exp\bigg\{\frac{-1}{2\sigma_c} \left[\mathcal{M}_c(m_1,m_2)-\mathcal{M}_{ci}\right]^2 \\
         -\frac{-1}{2\sigma_{\eta}} \left[ \eta(m_1,m_2)-\eta_i \right]^2 \bigg\}
\end{multline}
where $i$ represents each intrisic grid point in an injection. $\sigma_c$ and $\sigma_{\eta}$ are the standard deviations for the gaussians centered at $\mathcal{M}_{ci}$ and $\eta_i$, respectively. 

For the extrinsic parameters, we used the likelihood value calculated by Rapid PE for each of the extrinsic samples at every grid point along with the given priors to compute the CDF of the posterior up to the injected extrinsic value. 

\begin{figure*}[htbp]
\includegraphics[width=0.99\textwidth]{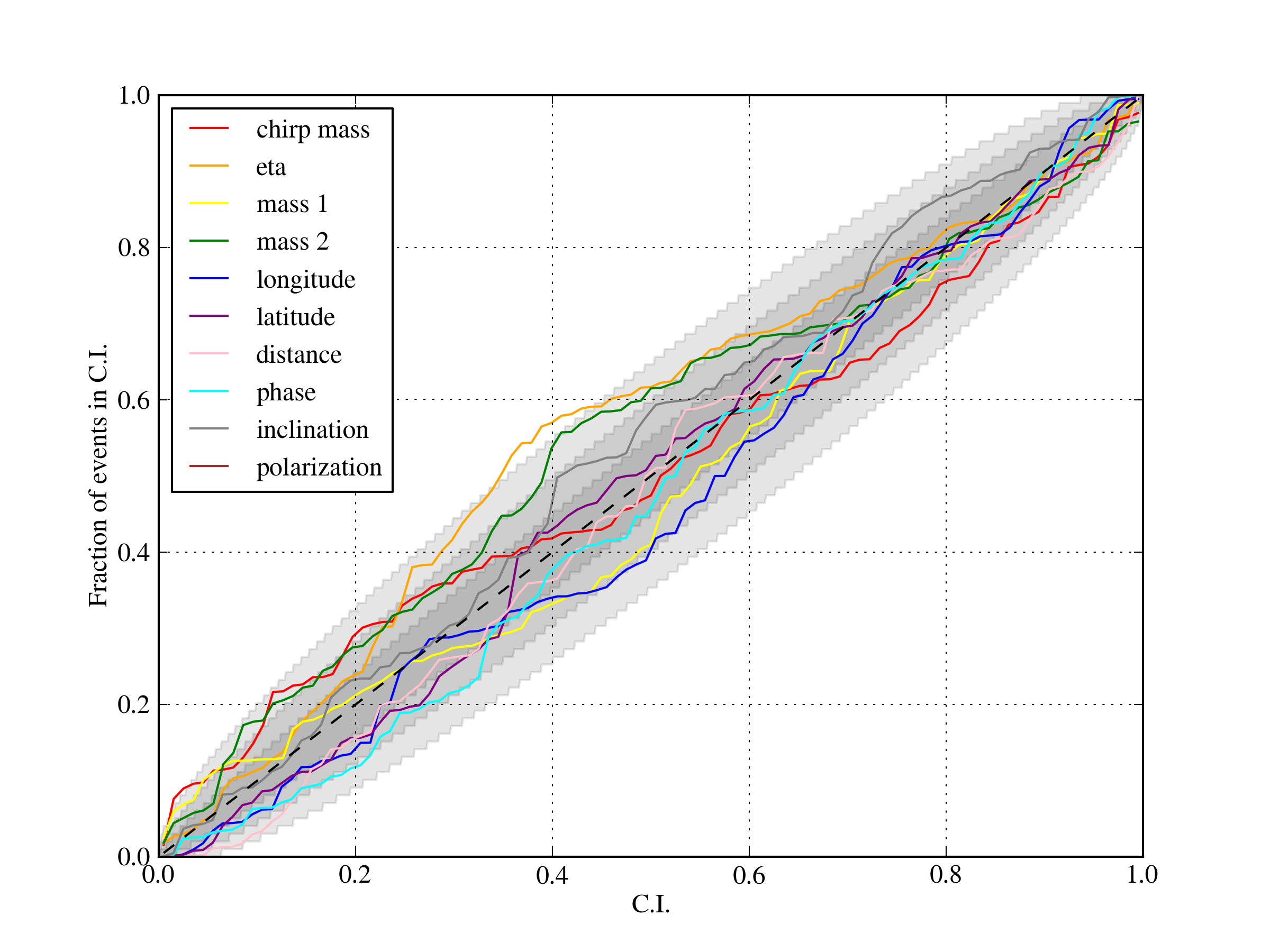}
\caption{\label{fig:pp_plot} Rapid PE results on 100 three-detector F2Y 2016 injections for the serial refinement method with three grid levels. The colored lines represent the fraction of events within a given confidence interval (C.I.) for intrinsic and extrinsic parameters.  The gray regions mark the cumulative $1-$, $2-$, and $3-\sigma$ confidence intervals in order of decreasing opacity.}
\end{figure*}

Figure \ref{fig:pp_plot} shows the result of this validation test, where the confidence interval is plotted against the fraction of injections within that confidence interval. The cumulative $1-$, $2-$, and $3-\sigma$ confidence intervals appear as ovals behind the P-P plots as in~\cite{Romero_Shaw_2020}. 

\section{Discussion and Conclusions}
\label{sec:discuss_conclude}
While the original version of Rapid PE introduced the method by which we can simplify and parallelize parameter estimation through restricting intrinsic parameters to a grid, it was not flexible enough to overcome search biases, especially in cases with higher spin. The improvements outlined in this paper tackle this issue through an adaptive grid procedure which refines the grid around points with more evidence to map out the posterior more completely. We demonstrate that in the case of significant search bias, adaptive grid refinements cover more of the posterior without sacrificing sampling efficiency compared to the Fisher matrix grid scheme of the original version and the initial grid of the prerefine method based on overlap computations alone. Moreover, we develop a rectilinear grid scheme extensible to spin to better parameterize a wide variety of events, including binary black holes. 

Quantitatively, all three methods have been shown to perform reasonably well in recovering the extrinsic parameters in producing similar numbers of effective samples. The main tension between the three methods is the trade off in computational efficiency, sampling efficiency, and adaptability. The Fisher and prerefine scheme require no serial steps, but the Fisher scheme has no clear extension to refinement. The prerefine scheme can suffer from incomplete coverage of the posterior if the true posterior support is not commensurate with expectations from the overlap. If the posterior is not adequately mapped, then ad-hoc follow up analyses would be necessary to correct the deficiency. Furthermore, when additional parameters are included in the intrinsic parameter set (for example, components of the compact object spins) refinement and point-pruning will become critical to ensure the posterior computation is completed in a prompt fashion. The Fisher matrix scheme performed adequately for the task of two dimensional parameter estimation, but is not flexible enough to handle search biases in a prompt manner. The serial refinement method is likely to be more accurate since it is actually mapping the likelihood surface, but each refinement level requires the evaluation of the new grid points on the likelihood surface. 

Complementary to this work, other methods to speed-up parameter estimation of compact binary coalescences from gravitational waves include fast ROQ~\cite{Morisaki:2020oqk}, machine learning~\cite{George:2017pmj}, accelerated waveform generation~\cite{Pratten:2020fqn,Cotesta:2020qhw,Lackey:2018zvw}, and GPUs~\cite{Talbot:2019okv,PhysRevD.99.084026}. RIFT is another extension of the original Fisher matrix method of Rapid PE which successfully employed GPUs to dramatically reduce the latency of parameter estimation~\cite{Lange:2018pyp,Lange:2017wki,PhysRevD.99.084026,LIGOScientific:2021djp, LIGOScientific:2021usb, LIGOScientific:2021qlt, LIGOScientific:2020ibl, LIGOScientific:2020ufj, LIGOScientific:2016kms}. Future work will include the use of GPUs to speed-up the marginalized likelihood calculations of the version of Rapid PE presented here, while retaining the benefits of using rectilinear grids with adaptive refinements to overcome search biases. 

Based on the validity tests illustrated by the P-P plots, we conclude that the updated version of Rapid PE presented in this work produces reliable intrinsic and extrinsic results for binary neutron star sources under a variety of conditions and signal to noise ratios. This work has shown that the gridded approach can map the posterior to a point where interpolation is a viable alternative to Markovian sampling and that search biases can be ameliorated through additional grid refinements. We expect that this method can be used to perform compact binary merger parameter estimation in a computationally efficient and prompt manner, benefitting low-latency electromagnetic follow-up of electromagnetically active sources.

\section{Acknowledgements}
\label{sec:acknowledgements}
This work was supported by the National Science Foundation awards PHY-1912649, PHY-1700765, and PHY-1626190. We are grateful for computational resources provided by the Leonard E Parker Center for Gravitation, Cosmology, and Astrophysics at the University of Wisconsin-Milwaukee and the LIGO Laboratory at the California Institute of Technology. We would like to thank Koh Ueno and Jolien Creighton for greatly beneficial discussions regarding the P-P and posterior plots. We would also like to thank Duncan Meacher, Heather Fong, Soichiro Morisaki, Daniel Wysocki, and Ignacio Magaña for their useful advice.

\bibliography{rapid_pe}
\end{document}